\documentstyle[sprocl]{article}
\def\be{\begin{equation}}
\def\ee{\end{equation}}
\def\bea{\begin{eqnarray}}
\def\eea{\end{eqnarray}}

\begin{document}
\title{DETECTING NEUTRINOS FROM AGNS AND \break TOPOLOGICAL DEFECTS 
WITH NEUTRINO TELESCOPES \footnote{Talk presented by I.\ Sarcevic.}}
\vskip -0.4true in
\author{Raj Gandhi,$^\ast$ Chris Quigg,$^\dagger$ Mary Hall
Reno,$^\ddagger$ and Ina Sarcevic$^\star$}
\baselineskip=12pt 
\address{{$^\ast$Mehta Research Institute, Allahabad 211002, India}\\ 
{$^\dagger$Fermi National Accelerator Laboratory, Batavia, IL 60510 USA}\\ 
{$^\ddagger$Department of Physics \& Astronomy, University of Iowa, Iowa 
City, IA 52242}\\ 
{$^\star$Department of Physics,
University of Arizona, Tucson, AZ 85721}}

\maketitle\abstracts{
We evaluate neutrino-nucleon cross section for
energies up to $10^{21}\hbox{ eV}$ in light of new information on the 
small-$x$ behavior of parton distributions.  We give predictions for large 
underground neutrino telescope event rates for ultrahigh-energy neutrinos 
from  Active Galactic Nuclei and from the decay of 
topological defects formed in the early Universe.  
}
\vskip -0.2true in

Active Galactic Nuclei (AGNs) are the most powerful 
sources of high-energy gamma rays.  If these gamma rays originate in 
the decay of $\pi^{0}$, 
then AGNs may also be prodigious sources
of high-energy neutrinos.  
Neutrinos are undeflected by magnetic fields and have 
long interaction lengths, so they may
potentially provide valuable information 
about astrophysical sources.  Gammas, 
on the other hand, are absorbed by a few hundred grams of material.  
As underground neutrino telescopes  
achieve larger instrumental areas, prospects for measuring fluxes from 
AGNs become realistic.  

The diffuse flux of AGN neutrinos, 
summed over all sources, is isotropic, so
the event rate is $A \int dE_\nu P_\mu(E_\nu,E_\mu^{\rm min}) 
S(E_\nu){dN_\nu/dE_\nu}$,
given a neutrino spectrum $dN_\nu/dE_\nu$ and detector
area $A$. 
Attenuation of neutrinos in the Earth, 
described by a shadowing 
factor $S(E_\nu)$, depends on the $\nu_\mu N$ cross section through the 
neutrino interaction length, while 
the probability 
that the neutrino
converts to a muon that arrives at the detector
with $E_\mu$ larger than the threshold energy $E_\mu^{\rm min}$, 
$P_\mu(E_\nu,E_\mu^{\rm min})$ 
is 
directly proportional to the charged-current cross section.  

Here we present predictions of
event rates for several models of the AGN  
neutrino flux.\cite{stecker}  
We also compare the predicted rates with the atmospheric neutrino 
background (ATM).\cite{volkova}  
These rates reflect a new calculation \cite{Gqrs} of the
neutrino-nucleon cross section that incorporates recent results from
the HERA $ep$ collider.\cite{He}

The classic signal for cosmic neutrinos is energetic muons produced in 
charged-current interactions of neutrinos with nucleons.  
To reduce the background from muons produced
in the atmosphere, we consider 
upward-going muons produced in and below the detector in $\nu_\mu N$ and 
$\bar\nu_\mu N$ interactions.  
We also give predictions for downward-moving 
(contained) muon event rates 
due to $\bar\nu_e e$ interactions in the
PeV range and for neutrinos produced in the collapse of 
topological defects.  

In Table \ref{upward} we show the event rates for a detector with $A=0.1\hbox{ 
km}^2$ for $E_\mu^{\rm min}=1\hbox{ TeV}$ and $10\hbox{ TeV}$.
The CTEQ--DIS rates are representative of the new generation of 
structure functions.\cite{CTEQ}  The older rates derived from the EHLQ 
structure functions are given for comparison.\cite{Rq}  If the most 
optimistic flux predictions are accurate, the observation of AGNs by 
neutrino telescopes is imminent.
\begin{table}[t!]
\caption{
Number of upward $\mu+\bar{\mu}$ events
per year per steradian for $A=0.1$ km$^2$.} 
\begin{center}
\begin{tabular}{ccccc}
\hline
\raisebox{-1.5ex}{Flux} & \multicolumn{2}{c} { $ E_\mu^{\rm min}=1\hbox{ TeV}$} 
& 
\multicolumn{2}{c}{ $ E_\mu^{\rm min}=10\hbox{ TeV}$} \\ 
 & EHLQ & CTEQ--DIS & EHLQ & CTEQ--DIS \\ \hline
AGN--SS \cite{stecker} & 82 & 92 & 46 & 51 \\ 
AGN--NMB \cite{stecker} & 100 & 111  & 31 & 34 \\ 
AGN--SP \cite{stecker} & 2660 & 2960 & 760 & 843  \\ 
ATM \cite{volkova}& 126 & 141 & 3 & 3 \\ \hline
\end{tabular}
\end{center}\label{upward}
\vskip -0.1 true in
\end{table}

Only in the neighborhood of $E_\nu=6.3\hbox{ PeV}$, where the $W$-boson is 
produced as a $\bar{\nu}_e e$ resonance, are electron targets 
important.  
The contained event rate for resonant $W$ production is
${(10/18)} V_{\rm eff}  N_A
\int dE_{\bar{\nu}} \sigma_{\bar{\nu}e}(E_\nu) 
S(E_{\bar{\nu}}){dN/dE_{\bar{\nu}}}$.
We show event rates for downward resonant $W$-boson production in Table 
\ref{electron}.  (The Earth is opaque to upward-going 
$\bar{\nu}_{e}$s at resonance.)
\begin{table}[b!]
\caption{
$\bar\nu_e e\rightarrow W^-$ events per 
year per steradian for a detector with effective volume 
1~km$^3$ and the downward (upward) background
from $(\nu_\mu,\bar\nu_\mu) N$ interactions above 3 PeV.}
\begin{center}
\begin{tabular}{ccc}
\hline
Mode & AGN--SS & AGN--SP \\ \hline
$W\rightarrow \bar{\nu}_\mu \mu$ & 6 & 3 \\ 
$W\rightarrow {\rm hadrons}$ & 41 & 19 \\ \hline
$(\nu_\mu,\bar\nu_\mu)N$ CC  & 33 (7) & 19 (4) \\ 
$(\nu_\mu,\bar\nu_\mu)N$ NC  & 13 (3) & 7 (1) \\ \hline
\end{tabular}
\end{center}\label{electron}
\end{table}
We note that a 1-km$^3$ detector with energy threshold in the PeV range 
would be suitable for 
detecting resonant $\bar\nu_e e\rightarrow W$ events,
though the $\nu_\mu N$ background is not negligible.

Another possible source of UHE neutrinos is
topological defects such as monopoles, cosmic strings, and
domain walls, which might have been formed 
in symmetry-breaking phase transitions in the early Universe. When 
topological defects are destroyed by collapse or annihilation, the energy 
stored in them is released in the form of massive $X$-quanta  
of the fields that generated the defects. The $X$ particles can then 
decay into quarks, gluons, leptons, and such, that eventually materialize 
into energetic neutrinos and other particles. 

Table \ref{TDrates} shows rates induced by the neutrino flux from the collapse 
of cosmic-string loops, in a model \cite{hill} that survives the Fr\'{e}jus 
bound \cite{frejus} at low energies.  
We take this flux as a plausible example to consider the 
sensitivity of a km$^{3}$ detector to fossil neutrinos from the 
collapse of topological defects.

\begin{table}[t]
\caption{ 
Downward $\mu^{+}+\mu^{-}$ events per steradian per year 
from $(\nu_{\mu},\bar{\nu}_{\mu})N$ interactions in
a detector with effective volume 1 km$^{3}$, for the 
BHS$_{p= 1.0}$ flux from topological defects.} 
\begin{center}
\begin{tabular}{ccc} \hline
\raisebox{-1.5ex}{Parton Distributions} & \multicolumn{2}{c}{$E_{\mu}^{\
mathrm{min}}$} \\ 
 & $10^7\hbox{ GeV}$ & $10^8\hbox{ GeV}$ \\ \hline
 CTEQ--DIS & 10 & 6 \\
 CTEQ--DLA & 8  & 4 \\
 MRS D\_ & 12 & 8 \\
EHLQ & 6  & 3 \\ \hline
\end{tabular}
\end{center}\label{TDrates}
\end{table}

For our nominal set (CTEQ-DIS) of parton distributions, the BHS$_{p = 
1.0}$ flus leads to 10 events per steradian per year with 
$E_{\mu}>10^{7}\hbox{ GeV}$, far larger than the rate expected from 
``conventional'' pion photoproduction on the cosmic microwave 
background.  This is an attractive 
target for a 1-km$^{3}$ detector, and raises the possibility 
that even a 0.1-km$^{3}$ detector could see hints of the collapse of 
topological defects.  

\vskip -1.5true in

\end{document}